\documentclass[twocolumn,secnumarabic,amssymb, nobibnotes, aps, prd]{revtex4}

\usepackage{color}
\usepackage{amsmath}
\usepackage{graphicx}
\usepackage{amssymb}
\usepackage{epstopdf}
\usepackage{slashed}

\begin{document}

\title{Dynamical violation of scale invariance and the dilaton in a cold Fermi gas}

\author{Gordon W. Semenoff and Fei Zhou }
\affiliation{
 Department of Physics and
Astronomy, University of British Columbia,
6224 Agricultural Road, 
Vancouver, British Columbia, Canada V6T 1Z1 }

\begin{abstract}
We use the large N approximation to find an exotic phase of a cold, two-dimensional, 
N-component Fermi gas which 
exhibits dynamically broken approximate scale symmetry.   
 We identify a particular weakly damped  
 collective excitation  as the dilaton,
the pseudo-Goldstone boson associated with the
broken approximate scale symmetry.   We argue that the symmetry breaking phase is stable for
a range of parameters of the theory and there is a fluctuation induced first order
  quantum  phase transition between the normal
and the scale symmetry breaking phases which can be driven by tuning the chemical potential.
We find that the compressibility of the gas at its lower critical density is anomalously large,
2N times that of a perfect gas at the same density.
\end{abstract}

\maketitle

 There has been some interest in approximately scale invariant relativistic quantum field theories where the scale symmetry
 can be spontaneously broken, resulting in the appearance of a light dilaton as a pseudo-Goldstone boson.  This idea has been a theme in
 walking technicolor theories for dynamical breaking of the symmetry in the standard model  \cite{Appelquist:2010gy}-\cite{Hashimoto:2010nw} as
  well as in the idea that the standard model Higgs field could itself be a light dilaton \cite{Goldberger:2008zz}-\cite{Grinstein:2011dq}.
 It also appears in the tricritical O(N) model  in three dimensions  where the large N limit has the Bardeen-Moshe-Bander 
 phase \cite{Bardeen:1983rv} which has spontaneously broken scale invariance and a massless dilaton. The latter  
is unstable when the infinite N limit is relaxed. The dilaton becomes a tachyon \cite{Omid:2016jve}.  
 It is interesting to ask whether there are other physically plausible contexts where approximate scale symmetry
 is dynamically broken, the symmetry breaking phase is stable 
 and where  theoretical analysis would be under good analytic control.  
 In the following we shall point out that the large N limit of a cold, non-relativistic, 
 N-component, two-dimensional Fermi gas with an
 attractive delta-function two-body interaction could
 be such a system. We shall study its quantum critical behaviour 
at the phase transition between a phase where it acquires a density of particles, and in which the density itself acts as an order parameter,
 and a phase where the density goes to zero.  This phase boundary becomes a line of
 first order phase transitions. This model exhibits an example of the Coleman-Weinberg
 mechanism for dynamical symmetry breaking \cite{Coleman}. What is more, it has
approximate scale symmetry and we shall identify the pseudo-goldstone mode corresponding to the dilaton. 

 Normally, a cold  Fermi gas with an attractive interaction would
 be expected to be a superfluid with a condensate of Cooper pairs.  Indeed, this should be
 the fate of the  system that we discuss.  However, we shall find that, at large N,  the pairing instability is exceedingly weak, with
 the Cooper pair binding energy $m_B$ exponentially small in N and also exponentially smaller than the 
 dynamically generated scale, $\rho_{\rm crit.}$, of the density, $m_B\sim e^{-2 N}\rho_{\rm crit}$ \cite{largeN}. 
(We use units  where $\hbar=1=2m$.)
  
  As an open system, where Fermions are allowed to flow in and out
 to a reservoir, the cold Fermi gas is described by the non-relativistic quantum field theory with 
 (imaginary time) action,  $S=\int dtd^2x{\mathcal L}$, with
 \begin{align}
{\mathcal L}=
\psi_a^*\dot\psi_a+\vec\nabla\psi_a^*\cdot\vec\nabla\psi_a  
 -\frac{g}{2{\rm N}}(\psi_a^*\psi_a)^2-\mu \psi_a^*\psi_a
\label{lagrangian_density}
\end{align}
and the grand canonical thermodynamic potential defined by the functional integral
\begin{equation}\label{pathintegral0}
 e^{-{\mathcal V}\Phi}=\int [d\psi_a(x) d\psi^*_a(x)] ~e^{-S[\psi,\psi^*] }
\end{equation}
where ${\mathcal V}$ is the volume of space-time, $\psi_a(x)$ are anti-commuting functions, 
 the paired indices $a$ run from 1 to N and are implicitly summed where they appear in the Lagrangian density
 and $\mu$ is the chemical potential.  
In the classical theory described by (\ref{lagrangian_density}), 
the chemical potential  is the only parameter  with  a non-zero  scaling dimension. It has dimension of
energy or 1/distance$^2$.   Setting it to zero would yield a scale invariant classical field theory.  However, 
it is well-known that quantization of the theory  described by (\ref{lagrangian_density}) and (\ref{pathintegral0}) breaks the scale invariance 
with a scale anomaly \cite{anom0}-\cite{an3}. 
The coupling constant $g$ obtains a beta-function \cite{Bergman:1991hf} 
from the renormalization of a logarithmic ultraviolet divergence which occurs
in Fermion-Fermion two-body scattering,  
\begin{align}\label{beta}
\beta(g(\Lambda^2)) = \Lambda^2 \frac{d}{d\Lambda^2}g(\Lambda^2)=-\frac{g^2}{8\pi{\rm  N}}+{\mathcal O}(\tfrac{1}{\rm N}^2) 
\end{align}
 and it becomes a scale-dependent running coupling
 \begin{align}
\frac{1}{g(\Lambda^2_1)}-\frac{1}{g(\Lambda^2_2)}=\frac{1}{8\pi {\rm  N}}\ln\frac{\Lambda^2_1}{\Lambda^2_2}
+{\mathcal O}(\tfrac{1}{{\rm N}^2})
\label{beta1}
\end{align}
where $\Lambda_i$ are scales with dimensions of 1/distance. We will call the ultraviolet cutoff $\Lambda$.   
The beta function is  small and the scale dependence is weak  in the large N  limit where N$\to\infty$ with $g$ fixed.  
In the strict infinite N limit, the model is scale invariant.  In the following, we shall study this model, both in the scale invariant
infinite N limit and at large but finite N where the scale symmetry is weakly broken.

At the infinite N limit, we shall find a state of this system which  can have a non-zero density per species, 
\begin{equation}\label{rho}
\rho = \frac{1}{{\rm  N}}\sum_{a=1}^{\rm  N}<\psi_a^\dagger(x)\psi_a(x)>
\end{equation}
when the coupling constant
is tuned to a strong coupling fixed point and the chemical potential is simultaneously tuned to zero.  
At large but finite N, we shall find that this
phase remains stable and that, within a wide band of renormalization group trajectories, 
 the running-coupling self-tunes to the fixed point.   

 We shall find that the large N limit of the theory described by (\ref{lagrangian_density}) has
 a thermodynamic potential
\begin{align}\label{potential}
\Phi = {\rm  N} ~\left[ g\frac{\rho^2}{2}- \frac{(\mu+g\rho)^2}{8\pi}\right]+\ldots
\end{align}
In this simple equation, the first term is the mean field potential energy  and the second term is the
Fermi pressure with the mean field effective chemical potential $\mu+g\rho$. 
The ellipses  denote corrections to the large N limit, which we shall ignore for the moment. 
This potential should be minimized as a function
of  $\rho$ while holding $\mu$ at a fixed value.  
The value of $\rho$ at the minimum is the physical value of the density. 
There is a solution only if $0<g\leq 4\pi$ and the result is 
an equation which relates the density and the chemical potential, 
\begin{align}\label{mu}
\mu=(4\pi-g)\rho +\ldots ~~,~~g=4\pi-\frac{\mu}{\rho}
\end{align}
and, upon substituting into (\ref{potential}),  determines the pressure
\begin{align}\label{pressure1}
P = {\rm  N}\frac{\rho^2}{2} \left(4\pi - g \right)+\ldots
\end{align}
In Eq.\  (\ref{mu}) we see that there are two ways to set the chemical potential to zero.  One is to 
put the density to zero, so that there are no particles.  The other is to  tune the coupling constant 
so that $g=4\pi$ which we could do while keeping $\rho$ non-zero.  
In this case, both the chemical potential and the pressure vanish for any value of the
density.  The scale symmetry is spontaneously broken with the density playing
the role of the order parameter. 
The compressibility  
\begin{align}\label{kappa}
\kappa = \frac{1}{(N\rho)^2} \frac{d(N\rho)}{d\mu} 
=\kappa_0\frac{4\pi}{4\pi-g}
\end{align}
 diverges and the critical point (where $\kappa_0$ is the compressibility of a perfect gas at the same density $\kappa_0= \frac{1}{N\rho^2}\frac{1}{4\pi}$). 
\begin{figure}
\includegraphics[scale=.23]{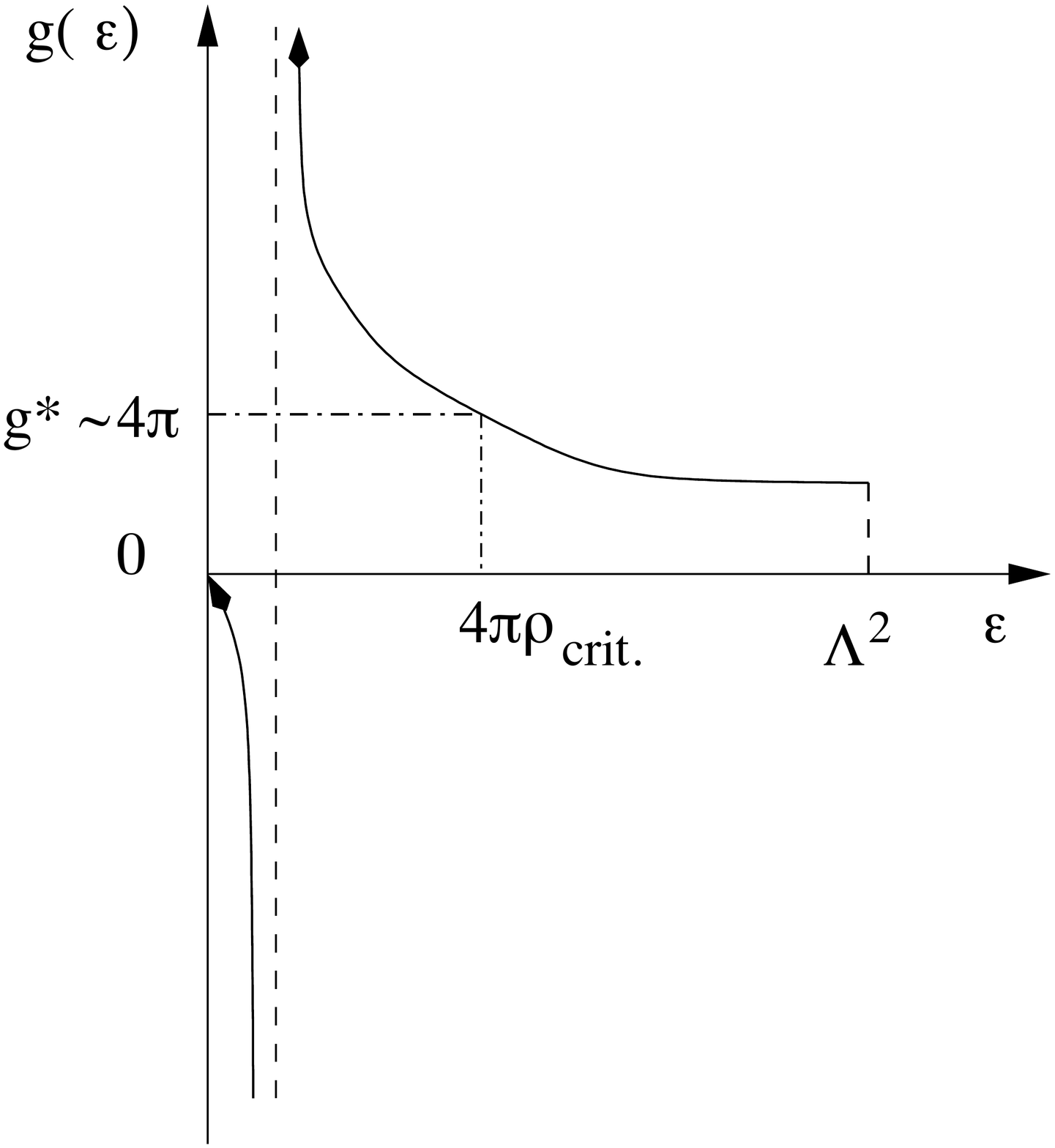}\\
\begin{caption} {  The flow of  $g $ as  the
 energy scale $\epsilon$  decreases. 
The critical density is the scale at which the running coupling reaches $g^*$, i.e. $g(4\pi\rho_{\rm crit.})=g^*$. 
See Eqs.\  (\ref{rhocrit1}) and (\ref{gstar}). 
The dashed line is the position of the Landau
pole which is exponentially smaller than $\rho$ in
the large N limit (  Eq.\ (\ref{landau_pole})).
    \label{flow}  
}\end{caption}
 \end{figure}  
 
We shall find that a non-propagating, non-oscillating collective mode emerges at this critical point.  We call this mode  
the dilaton.  It is a close relative  of the
breathing mode of cold atoms in a harmonic trap which is predicted by  a dynamical conformal algebra that appears there  \cite{Pitaevskii}.  
In that case the frequency of the breathing
is related to the harmonic potential of the trap. We can view our system as the limit where the trapping potential is removed (while keeping the
density fixed and simultaneously tuning the chemical potential to zero and the two-body coupling to the fixed point)
so that the frequency of oscillation of the breathing mode goes to zero. 

A crucial question  is that of stability of the phase with finite density per Fermion species, particularly when  scale symmetry violation is
turned on by relaxing the infinite N limit.   
We will find
that including order 1/N corrections helps in this regard.    When N is not strictly
infinite, the coupling $g$ has the beta-function in Eq.\  (\ref{beta}) which results in a order 1/N
violation of scale invariance.  The renormalization group flow of the
 coupling constant will make it a  
 function of the density and the equilibrium density will be determined by the equation 
 $g(4\pi\rho)=g^*$ where $g^*=4\pi+{\mathcal O}(1/{\rm N})$.  As illustrated in Fig.\ \ref{flow},
 this equation always has a solution on the attractive branch of the running coupling.

\begin{figure}
\includegraphics[scale=.20]{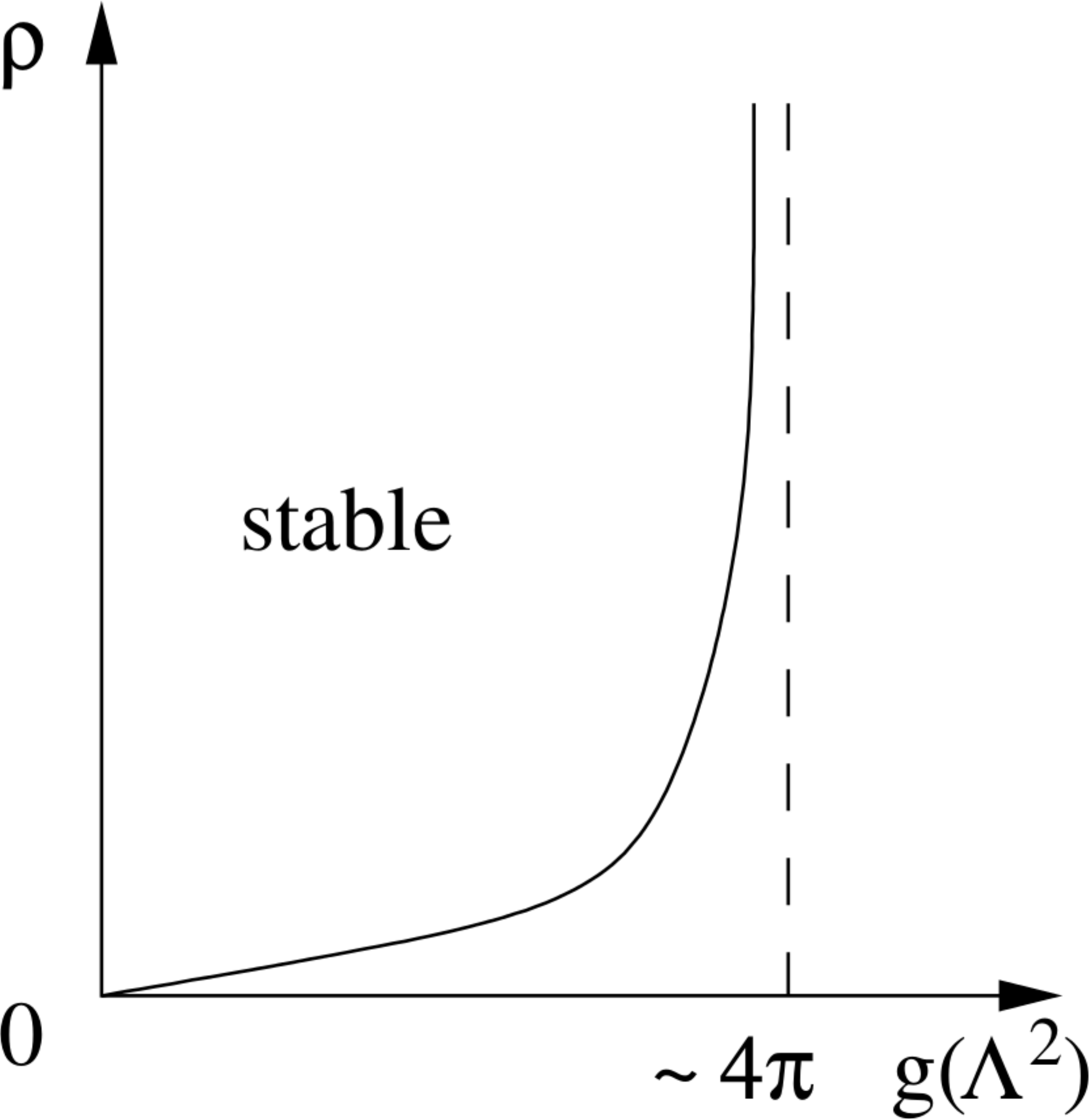}\\
\begin{caption} {   The solid line is the critical density $\rho_{\rm crit.}$ which forms the 
phase boundary at a line of first order phase transitions.  The  homogenous Fermi liquid phase
  is above the line and what is likely a clumped inhomogeneous phase is below it. The figure is schematic.  The
space between the lower critical density and $\rho=0$ axis is exaggerated.  The phase in this region is likely a mixed phase with clumps of fermions and the first order transition is very similar to a liguid-gas transition. 
See  also Fig.\ \ref{phase2}.
\label{phase}  
}\end{caption}
 \end{figure}  
 
 To find the grand canonical potential to the next-to-leading order in 1/N, 
it is convenient to begin with the action  $\tilde S = \int dtd^2x\tilde{\mathcal L} $
\begin{align}
\tilde{\mathcal L} = \psi_a^*\dot\psi_a+\vec\nabla\psi_a^*\cdot\vec\nabla\psi_a
 -(\mu+ g\sigma)\psi_a^*\psi_a
  +\frac{{\rm  N}g\sigma^2}{2}  
\end{align}
and  the functional integral
\begin{equation}\label{pathintegral}
 e^{-{\mathcal V}\Phi}=\int [d\psi_a(x) d\psi^*_a(x)d\sigma(x)] ~e^{-\tilde S[\psi,\psi^*,\sigma] }
\end{equation}
The Hubbard-Stratonovich $\sigma$-field is  related to the Fermion density per species, for example
$
\left< \sigma(x)\right>=   \rho $ and $<\sigma\sigma>_C=<\rho\rho>_C-1/g$. 
If we first do the Gaussian functional integral
over  $\sigma(x)$,  we recover the purely Fermionic theory  (\ref{lagrangian_density}).  
If, instead of integrating $\sigma(x)$, we first perform 
the Gaussian integral over $\psi_a(x)$, we obtain the effective action for the $\sigma(x)$-field
\begin{equation}
S_{\rm eff}=  \int dtd^2x \frac{{\rm  N}g\sigma^2}{2}  
-{\rm  N}{\rm Tr}\ln(\partial_t-\nabla^2-(\mu+g\sigma))
\end{equation}
This effective action is proportional to N and the dynamics of $\sigma(x)$ are therefore weakly coupled 
and semi-classical when N is large.  
The remaining functional integration over the variable $\sigma(x)$ can be computed  in the 
 saddle point approximation.  
To proceed, we
 denote $\sigma(x)=\rho+\delta\sigma(x)$.   
  We expand $S_{\rm eff}$ to quadratic order in $\delta\sigma(x)$,  we drop the linear terms,  
 and we do the Gaussian integral over $\delta\sigma(x)$ \cite{Jackiw:1974}.  The result is
 the potential density
 \begin{align}
 \Phi =& \frac{{\rm  N}g\rho^2}{2}  
-\frac{\rm  N}{\mathcal V}{\rm Tr}\ln D^{-1} 
+\frac{1}{2\mathcal V}{\rm Tr}\ln \Delta^{-1}
+\ldots
\label{effective_action}\end{align}
where  the ellipses denote terms of order 1/N and 
\begin{align}
&\Delta^{-1}(x,y)=N\left[g\delta(x,y)+g^2\Pi(x,y)\right]\label{PI}
\\
&
\Pi(x,y)=D(x,y)D(y,x)
\label{S}\\
&
D(x,y)=(x|\frac{1}{\partial_t-\nabla^2-(\mu+g\rho)} |y) \label{D}
 \end{align}
 The density $\rho$ is to be determined so that it minimizes $\Phi$.
 If we  keep only the leading
 order in N (the first two terms in (\ref{effective_action})) and cancel divergent terms which are quadratic and quartic in the cutoff 
 with counter-terms (there are no logarithmic divergences 
 at this order), the potential  is as in Eq.\  (\ref{potential}) whose
 implications we have already discussed in the text following Eq.\  (\ref{potential}).
 
 Before we go on to discuss the next-to-leading order, let us consider the collective behaviour
 at the leading order at large N.   
 The $\sigma$-field propagator is $\Delta$ in (\ref{PI}).  We search for collective modes by studying the singularities of $\Delta(i\omega,k)$
 analytically continued to real frequency $i\omega\to \omega$. 
  The elementary one-loop integral yields
\begin{align}
\Pi(\omega,\vec k)= &  \frac{ (\omega+{k}^2+i\epsilon) \sqrt{1-    \frac{ 4(\mu+g\rho){k}^2}  {(\omega+{k}^2+i\epsilon)^2}  } -k^2} {8\pi {k}^2} \nonumber  \\
&+\frac{ (-\omega+{k}^2-i\epsilon) \sqrt{1-    \frac{ 4(\mu+g\rho){k}^2}  {(-\omega+{k}^2-i\epsilon)^2}  } -k^2} {8\pi {k}^2}   \nonumber 
\end{align}
 When the interaction is attractive and subcritical, that is, when $0\leq g<4\pi $, the equation $\Delta^{-1}=N[g+g^2\Pi(\omega,\vec k)]=0$ has no real solution for $\omega$.  This is 
consistent with the fact that zero sound is strongly damped in a Fermi liquid with 
an attractive interaction. 
 However,   
if we tune the coupling to $g=4\pi$,  $\Delta^{-1}=N[g+g^2\Pi(\omega,\vec k)]$  has a zero, and $\Delta$ a pole,  at $\omega=0$.  
Near the critical point, and in the $\omega<<p\sqrt{4\pi\rho}<<4\pi\rho$ regime, 
\begin{align}\label{ss}
\Delta(\omega,k)=\frac{\frac{4\pi}{{\rm N}g^2}\sqrt{\frac{\rho}{\pi}}|\vec k|}{\Gamma(k)-i\omega}
~,~
\Gamma(k)=\left(\frac{4\pi}{g}-1\right)  \sqrt{\frac{\rho}{\pi }}|\vec k| 
\end{align}
This collective mode is the dilaton.  It is  damped unless  $g=4\pi$, where the  scale symmetry
of the underlying model becomes exact.  At that point, $\Gamma=0$ and the pole is at $\omega=0$. 
 
  To proceed to the next order of the 1/N expansion, consider the integral which evaluates the last term on the right-hand-side of
Eq.\  (\ref{effective_action}).  
 \begin{align}
& \frac{1}{2\mathcal V}{\rm Tr}\ln\Delta^{-1}=\frac{1}{2}  \int \frac{d\omega}{2\pi}\int_0^{\Lambda^2}\frac{dk^2}{4\pi}\ln \Delta^{-1}( \omega,  k^2)  
\nonumber
 \\
&= \left[ -\frac{g^2 (\mu+g\rho)^2}{256\pi^3}\ln
\frac{\Lambda^2}{\mu+g \rho}
+\frac{(\mu+ g\rho)^2}{8\pi}\frac{\varphi(g)}{4\pi}+\ldots\right]
\end{align}
where we have isolated the quartic, quadratic and logarithmically divergent terms, 
dropped the quartic and quadratically divergent ones, assuming that they are 
canceled by counter-terms that we would add to the original Lagrangian density and we name the
remaining finite integral  $\frac{( \mu+g\rho)^2}{4\pi}\frac{\varphi(g)}{4\pi}$. Since it is finite, its  $ \mu+g\rho$-dependence is
given by dimensional analysis.  Our results will not depend on the precise form of this function. 
The renormalization of this theory  is well-known and in the leading order it is confined to
coupling constant renormalization which leads to the beta function (\ref{beta}).    What is more, the logarithmic terms spoil the
large N expansion.  When evaluated on the solutions for $\rho$,  the logarithms are large and they compensate the small factors of $1/$N.  
This is a well-known phenomenon
which already occurs in the first example in the paper by Coleman and Weinberg \cite{Coleman} on dynamical symmetry breaking.  Its solution
is to use the renormalization group to re-sum the logarithmically singular terms to all orders.  The result in our case is simple: it is given by replacing  
  $g$ in the potential by the running coupling constant $g(\tilde\mu )$ at the scale of the effective chemical potential, $\tilde\mu=\mu+g\rho$. 
Then, the renormalization group improved potential at the next-to-leading order in the large N expansion has the form
 \begin{align}
\Phi &= {\rm N}  \left[ \frac{g(\tilde\mu )\rho^2}{2}  -\frac{\tilde\mu^2  } {8\pi}\left(1-\frac{\varphi(g(\tilde\mu ))}{4\pi\rm N}\right)  \right] 
\nonumber \\
&+g(\tilde\mu )\left(\rho-\frac{\tilde\mu }{4\pi}\right)^2\frac{g(\tilde\mu)^2}{16\pi {\rm N}}\ln\frac{\Lambda^2}{\tilde\mu }
 +\ldots 
\label{final} \end{align} 
 The cutoff dependence has not entirely canceled.  However, the  cutoff-dependent term is
equal to the square of the derivative by $\rho$ of the leading order terms and it and its derivative will vanish to the order
in 1/N to which we are working  when it is evaluated on the 
solution of the equation  for $\rho$.  
The equation  $\frac{\partial\Phi}{\partial\rho}=0$ yields 
 \begin{align}
\mu &= \rho\left[  
 4\pi -{g(\tilde\mu )} +\frac{\varphi(g(\tilde\mu )) }{\rm N}-\frac{2\pi \beta(g(\tilde\mu ))}{g(\tilde\mu )}\right] +\ldots
\label{rho} 
\\   \label{rho1} 
\tilde\mu &=\mu+g(\tilde\mu )\rho~,~\tilde\mu \frac{d}{d\tilde\mu }g(\tilde\mu )=\beta(g)=-\frac{g^2}{8\pi\rm N}+\ldots
\end{align}
Eqs.\  (\ref{rho}) and (\ref{rho1}) determine $\rho$,  and therefore $\tilde\mu$  as functions of $\mu$. 
Eq.\ (\ref{rho}) is identical to (\ref{mu}) with   1/N corrections. When $\mu=0$, there are
two solutions of (\ref{rho}) and (\ref{rho1}), either $\rho=0$,  or $\rho\neq 0$ with the latter determined so that the running coupling constant obeys the equation 
$0= 4\pi -{g(\tilde\mu )} +\frac{\varphi(g(\tilde\mu )) }{\rm N}-\frac{2\pi \beta(g(\tilde\mu ))}{g(\tilde\mu )}$
with $\tilde\mu =g(\tilde\mu )\rho$. Which of these is the physical solution is obtained by comparing their thermodynamic potentials, $\Phi$. We see that $\Phi=0$ when $\rho=0$ and   $\Phi<0$ when $\rho$ is the other, non-zero solution.  Thus, when $\mu=0$, 
the solution with $\rho\neq 0$ is energetically preferred.   Now,
let us further decrease $\mu$ from zero to small negative values.  The phase with $\rho=0$ remains at $\Phi=0$ and the
phase with $\rho\neq0$ has $\Phi$ increasing and eventually coming to $\Phi=0$ to compete with the $\rho=0$ phase.  
This is the point of first order phase transition, where both phases have the same (zero) value of the thermodynamic potential.  By plugging the solution for $\rho$ into (\ref{final}), we see that $\Phi$ vanishes when the running coupling constant comes to the value $g^*$ which
obeys the equation  
\begin{align}\label{gcrit}
 4\pi-{g^*} +\frac{\varphi(g^*) }{\rm N}- \frac{4\pi \beta(g^*)}{g^*}+{\mathcal O}(1/{\rm N}^2)=0
 \end{align}
 This equation has solution
  \begin{align}
 g^*=4\pi\left( 1+\frac{\varphi(4\pi) }{4\pi\rm N}- \frac{\beta(4\pi)}{4\pi}+{\mathcal O}(1/{\rm N}^2)\right)
 \label{gstar}
   \end{align}
 and, the lower critical density is given by 
 \begin{align}\label{rhocrit1}
g(4\pi\rho_{\rm crit.})=g^* ~,~ \rho_{\rm crit.} = \frac{\Lambda^2}{4\pi} e^{-8\pi N\left( \frac{1}{g(\Lambda^2)}-\frac{1}{g^{*}}\right)}
\end{align}
  The pressure, is 
 \begin{align}P&=  {\rm N}\frac{\rho^2}{2} \left[  
   4\pi - {g( \tilde\mu ) } +\frac{ \varphi(g( \tilde\mu ) ) }{\rm N}+ \frac{ g( \tilde\mu )}{2N } \right] 
 +\ldots
\label{final1}\\
 \tilde\mu &=4\pi\rho+{\mathcal O}(1/{\rm N})
 \end{align}
  Eq.\ (\ref{final1}) is equal to the pressure in (\ref{pressure1}) plus $\frac{1}{\rm N}$ corrections.
 The pressure vanishes when $\rho=\rho_{\rm crit.}$.
 The critical chemical potential is of order 1/N and negative, 
 \begin{align}
 \mu_{\rm crit.}= 2\pi\rho_{\rm crit.}\frac{\beta(g^*)}{g^*}=-\frac{\pi }{\rm N}\rho_{\rm crit.}
 \end{align}  
  The compressibility   
  no longer diverges at the critical point,  
\begin{align}
\kappa = \kappa_0    \frac{4\pi}{g^*-g(\tilde\mu )+\frac{1}{2\rm N}g(\tilde\mu ) }
=\frac{  2N+\ln\frac{\rho}{\rho_{\rm crit.}}}{1+\ln\frac{\rho}{\rho_{\rm crit.}}}
\end{align}
where, as in (\ref{kappa}), $\kappa_0$ is the  compressibility of a perfect Fermi gas at the same density.  At the critical density, rather than diverging, the compressibility is anomalously large, $2$N times $\kappa_0$.
The damping constant of the dilaton at the critical density is obtained from the curvature of the potential, 
$ \frac{\partial^2\Phi}{\partial\rho^2}$ at $\rho=\rho_{\rm crit.}$, where it no longer goes to zero but it is suppressed by a factor of 1/N.  Eq.\  (\ref{ss}) is replaced by
\begin{align}\label{ss1}
\Delta(\omega,k)=\frac{\frac{1}{{\rm N}\pi}  \sqrt{4\pi \rho_{\rm crit.} }|\vec k|}{\Gamma-i\omega}
~,~
\Gamma=  \frac{1 }{{\rm N}}    \sqrt{4\pi \rho_{\rm crit.}   }|\vec k| 
\end{align}
The pressure and compressibility are positive, indicating stability of this phase
for values of $\mu$  larger than $\mu_{\rm crit.}$.  

 \begin{figure}
\includegraphics[scale=.34]{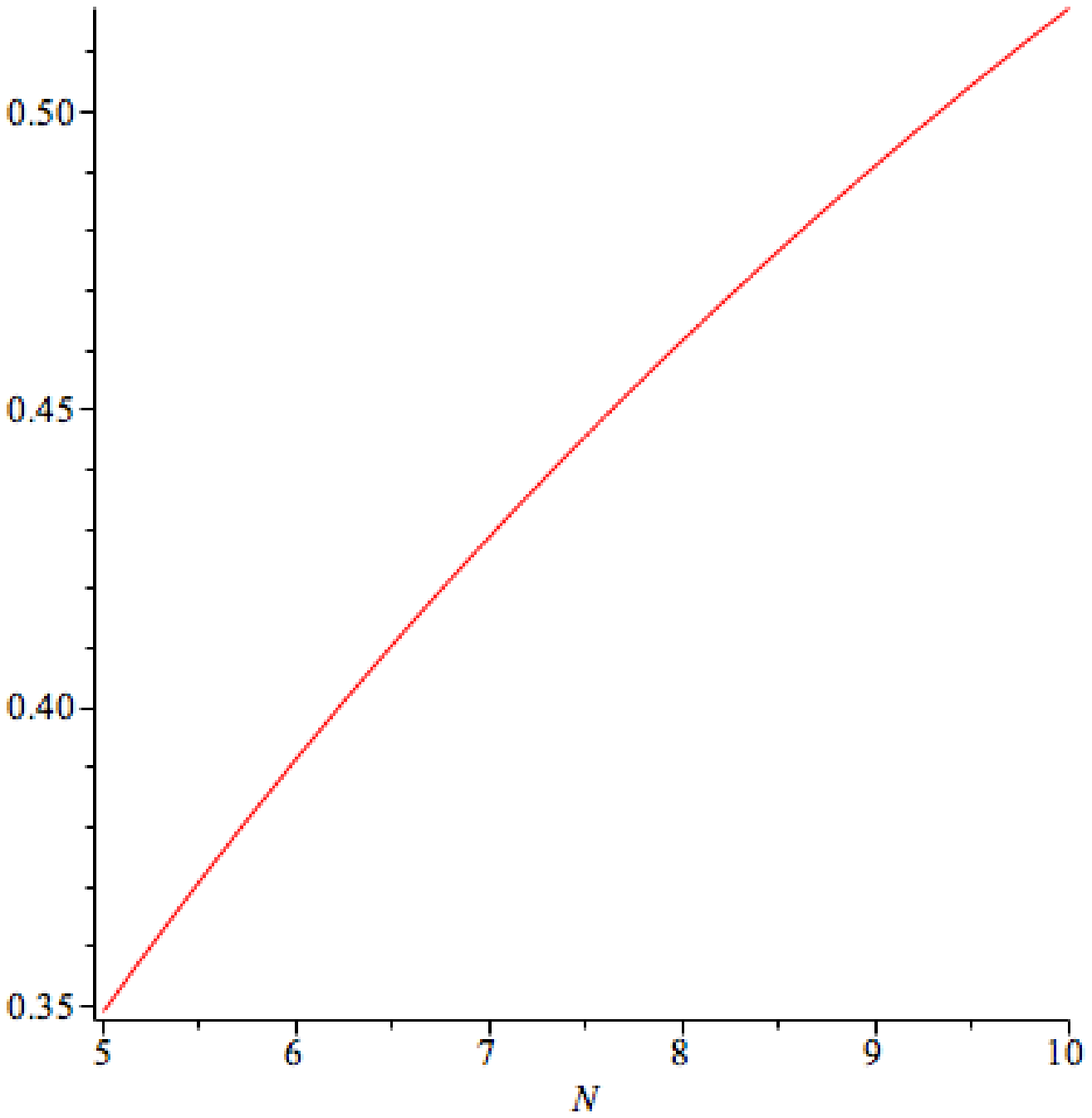}\\
\begin{caption} {   The  critcial value of $g(\Lambda^2)/g^*$ versus N,  with density  fixed to the 
physically reasonable value  $\rho =10^{-7}\Lambda^2$.\label{phase2}  
}\end{caption}
 \end{figure}  
  \begin{figure}
\includegraphics[scale=.36]{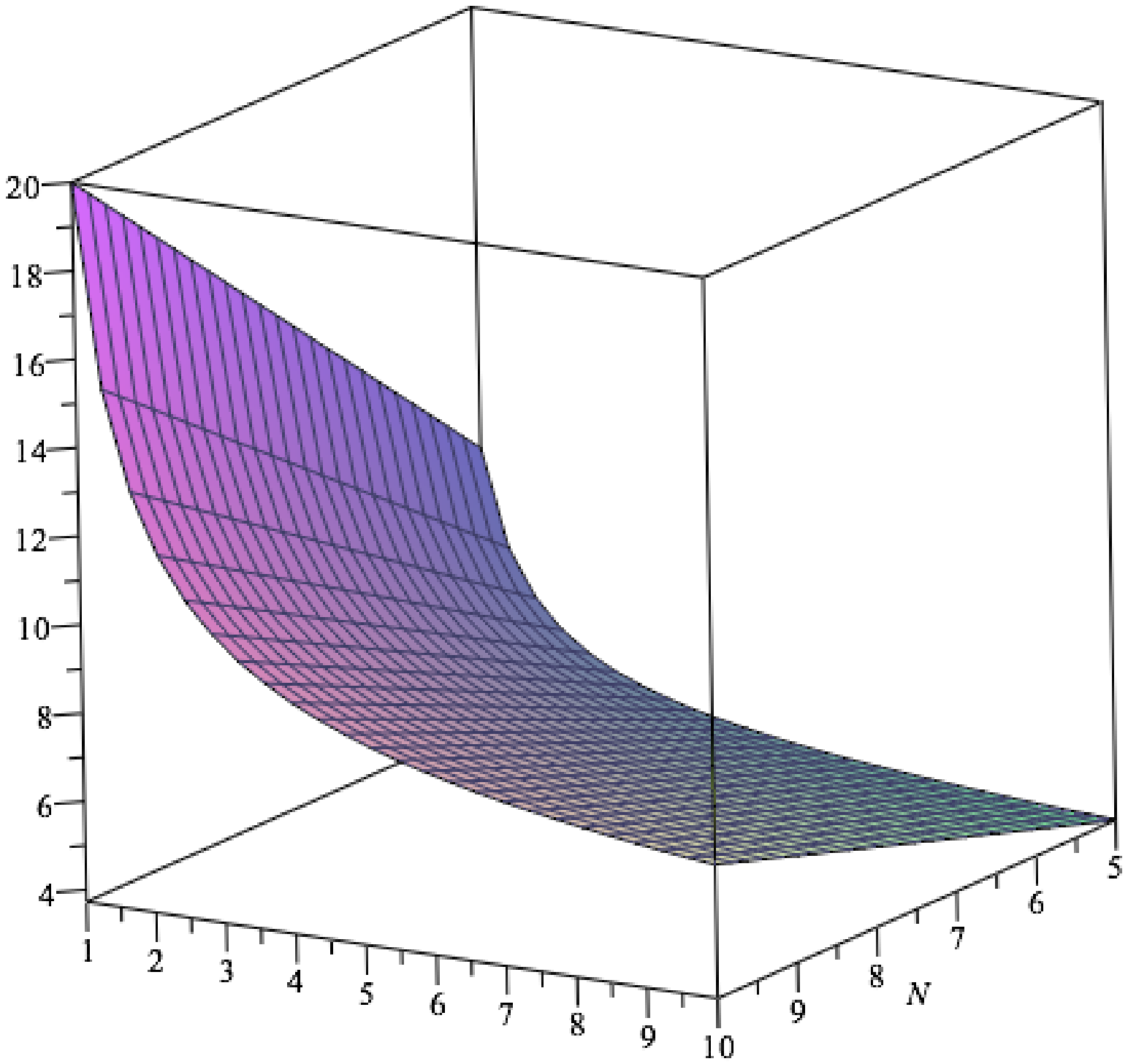}\\
\begin{caption} {   The compressibility $\kappa/\kappa_0$ is plotted on the vertical
axis versus  N on the $x$-axis and $\rho/\rho_{\rm crit.}$  on the y-axis.   \label{phase3}  
}\end{caption}
 \end{figure}  
  
  The first order behaviour arises from the fact that, as  seen in Fig.\ \ref{flow}, 
  the density is always larger than the scale of 
the  Landau pole which occurs at 
\begin{align}\label{landau_pole}
m_B= \Lambda^2 e^{- \frac{8\pi N}{g(\Lambda^2)}  } \sim  e^{-  2N} \rho_{\rm crit.}
\end{align}
This is the binding energy of a Cooper pair which appears in the Fermion-Fermion channel.
This weakly bound pair should Bose condense at zero temperature so that, in the strict mathematical sense,
the final state of this system should be a superfluid.  This Bose condensation is only possible if there is a density of
Fermions to pair and it should therefore aid the stability of the symmetry breaking phase by further lowering its free energy. 
However, the Cooper pair  binding energy is exceedingly small and
even a   small temperature  in the range
$m_B<k_BT<\rho$ or other small randomizing effects 
would destabilize the superfluidity but not the spontaneous density.

 Although the system that we have described is largely paradigmatic  it could conceivably 
 be experimentally realizable, for example,  
using Fermionic atoms  with high atomic or  nuclear spin  and  approximately spin independent attractive   interactions \footnote{As a result of the first order phase transition in large-N fermion gases
discussed here, we anticipate a phase separation phenomena
in a trap if the trap potential is slowly varying and the
local density near the centre is higher than the critical
one discussed in the article. Away from the centre of a trap with relatively
higher densities, the density profile along the radial direction
becomes sharply cliffed at the point where the
density decreases to the critical value; and beyond
that point, the fermions likely form a clumped liquid state
but spreading over a very narrow region.
In 2D traps, this implies that there
shall an interface which separates an inner disk
of stable fermions near the centre and a thin surrounding ring
of liquid at the outer boundary. This distinct singular
density profile can be the potential smoking gun of the
large N physics discussed in this Letter.}.  Such systems have been realized for N up to N=10 \cite{needs}, which should
be large enough to create a hierarchy of scales in Eq. (\ref{landau_pole}). 
In Fig.\  \ref{phase2} we plot the  upper critical value of the coupling $g(\Lambda^2)/g^*$ versus N  in the  interval $5\leq N\leq 10$ for a fixed and reasonable value of the density $\rho/\Lambda^2= 10^{-7}$.   The  compressibility is plotted versus the ratio of the density to the critical density for that range of N is plotted in Fig.\  \ref{phase3}.  The large enhancement of the compressibility, by a factor of 2N, at $\rho\to\rho_{\rm crit.}$ is clearly seen. The pressure and compressibility of the two-dimensional Fermi gas with attractive interactions, nonzero temperature and N=2 have been studied numerically in Ref.\  \cite{drut}. If such a study is extendable to the large N regime, it would be interesting as it would, for example, be possible to explore whether the phenomenon which we describe would be visible for small or intermediate values of N.

\begin{acknowledgments}
 This work is supported in part by NSERC Discovery grants 288179 and  1197908.  
 GWS thanks L.C.R.Wijewardhana for discussions and participation in an early state of this 
 work and Igor Herbut, Lev Lipatov and Arkady Vainshtein for  discussions.  
 \end{acknowledgments}

\end{document}